\documentclass{ifacconf}

\usepackage{natbib}            
\usepackage{graphicx,psfrag}          
\usepackage{amsmath}
\usepackage{graphicx}%
\usepackage{amsfonts,amsmath,amssymb}
\usepackage{mathrsfs,subfigure}

\newtheorem{assumption}[thm]{Assumption}

\def\cbb{\mathbb{C}}

\def\l1{\mathcal{L}^1}

\def\ipsdpoly0p{\overline{Q^{-1}(\mathfrak{Q}_{+}(n,\cbb))}}

\begin{document}

\begin{frontmatter}

\title{Analysis of Linear Quantum Optical Networks\thanksref{footnoteinfo}} 

\thanks[footnoteinfo]{This work was supported by the Australian Research Council.}

\author[First]{Ian R. Petersen}

\address[First]{School of
    Engineering and Information Technology, University of
    New South Wales at the Australian Defence Force Academy, Canberra,
    ACT 2600 Australia (email: i.r.petersen@gmail.com)}


\begin{abstract}
This paper is concerned with the analysis of linear quantum optical networks. It provides a systematic approach to the construction a model for a given quantum network in terms of a system of quantum stochastic differential equations. This corresponds to a classical state space model. The linear quantum optical networks under consideration consist of interconnections between optical cavities, optical squeezers, and beamsplitters. These models can then be used in the design of quantum feedback control systems for these networks. 
\end{abstract}

\end{frontmatter}

\section{Introduction}
This paper is concerned with the problem of network analysis for  linear
quantum optical networks. 
In recent years, there has been considerable interest in the modeling and feedback
control of
linear quantum systems; e.g., see
\cite{JNP1,NJP1,ShP5}.
Such linear quantum  
systems commonly arise in the area of quantum optics; e.g., see
\cite{GZ00,BR04}. Some recent papers have been concerned with the problem of realizing given quantum dynamics
 using physical
 components such as optical cavities, squeezers, beam-splitters, optical
amplifiers, and phase shifters; see \cite{PET08A,NUR10A}. 

This paper is concerned with the problem of constructing a dynamic model, in terms of quantum stochastic differential equations (QSDEs) (e.g., see \cite{JNP1}), for a general linear
quantum optical network consisting of an optical interconnection between optical cavities, squeezers and beam-splitters. This problem can be considered a quantum optical generalization of the classical electrical circuit analysis problem in which a state space model of the circuit is desired; e.g., see \cite{AV73,vVA74}. A systematic approach to the modelling of large quantum optical networks is  important as the construction of these networks is becoming feasible using technologies such as  quantum optical integrated circuits; e.g., see \cite{PCRYO06}. These QSDE models  can then be used in the design of a suitable quantum feedback controller for the network; e.g., see \cite{JNP1,NJP1}. This paper also describes how to construct alternative $(S,L,H)$ models for linear quantum optical networks; e.g., see \cite{GJ09}. These models can also be used for controller design or system simulation; e.g., see \cite{GJ09,PUJ2}. 

\section{Linear Quantum Systems}
In this section, we describe the general class of quantum systems under consideration; see also \cite{JNP1,GJ09,ZJ11}. 
We consider a
collection of $n$ independent quantum harmonic oscillators. Corresponding to this collection
of harmonic oscillators is a vector of  {\em annihilation operators}
$
a = \left[\begin{array}{cccc}a_1&a_2& \ldots & a_n
\end{array}\right]^T
$
on the underlying Hilbert space $\mathcal{H}$. 
The adjoint of the operator $a_i$ is
  denoted by $a_i^*$
and is referred to as a \emph{creation operator}. The operators
$a_i$ and $a_i^*$ are such that the following
 \emph{commutation
relations} are satisfied:
\begin{eqnarray}
\label{CCR2} 
\left[\left[\begin{array}{l}
      a\\a^\#\end{array}\right],\left[\begin{array}{l}
      a\\a^\#\end{array}\right]^\dagger\right]
&=&\left[\begin{array}{l} a\\a^\#\end{array}\right]
\left[\begin{array}{l} a\\a^\#\end{array}\right]^\dagger
- \left(\left[\begin{array}{l} a\\a^\#\end{array}\right]^\#
\left[\begin{array}{l} a\\a^\#\end{array}\right]^T\right)^T\nonumber \\
&=& \Theta
\end{eqnarray}
where $\Theta$ is a Hermitian commutation matrix of the form $\Theta = T J T^\dagger$ with $J = \left[\begin{array}{cc}I & 0\\
0 & -I\end{array}\right]$ and $T =  \Delta(T_1,T_2)$. Here $\Delta(T_1,T_2)$ denotes the matrix 
$\left[ \begin{array}{cc}
 T_1 & T_2 \\ T_2^\# & T_1^\#
     \end{array}  \right]$. Also, $^\dagger$ denotes the adjoint transpose of a vector of operators or the complex conjugate transpose of a complex matrix. In addition, $^\#$ denotes the adjoint of a vector of operators or the complex conjugate of a complex matrix. 

The quantum harmonic oscillators 
 are assumed to be
coupled to $m$ external independent quantum fields modelled by
bosonic  field annihilation operators $\mathcal{U}_1,
\mathcal{U}_2,\ldots,\mathcal{U}_m$. 
For each  field annihilation operator
 $\mathcal{U}_k$, there is a
corresponding  field creation operator
 $\mathcal{U}_k^*$, which is
 the operator adjoint of
$\mathcal{U}_k$. The field annihilation operators are also collected
 into a vector of
operators defined as follows:
$
 \mathcal{U}=\left[\begin{array}{cccc}
\mathcal{U}_1& \mathcal{U}_2& \ldots & \mathcal{U}_m
\end{array}\right]^T.
$
We also define a corresponding vector of output field operators $\mathcal{Y}$; e.g., see \cite{ZJ11}. The corresponding quantum white noise processes are defined so that 
$\mathcal{U}(t) = \int_0^tu(\tau) d\tau$ and $\mathcal{Y}(t) = \int_0^ty(\tau) d\tau$; e.g., see \cite{ZJ11}.

In order to describe the dynamics of a quantum linear system, we first specify the {\em
Hamiltonian operator} for the quantum system which is a Hermitian
operator on the underlying Hilbert space $\mathcal{H}$ of the form
\[
{\bf H} =
 \frac{1}{2}\left[\begin{array}{cc} a^\dagger &
       a^T\end{array}\right] M
\left[\begin{array}{c} a \\  a^\#\end{array}\right]
\]
where $ M$ is a Hermitian matrix of the
form
\begin{equation}
\label{tildeMN}
 M= \Delta( M_1, M_2).
\end{equation}
Also, we specify the {\em coupling operator vector} for the quantum
system to be a vector of  operators of the form
\[
L = \left[\begin{array}{cc} N_1 &  N_2 \end{array}\right]
\left[\begin{array}{c} a \\  a^\#\end{array}\right]
\]
where $ N_1 \in \mathbb{C}^{m\times n}$ and $ N_2 \in
\mathbb{C}^{m\times n}$. We can write
\[
\left[\begin{array}{c}L \\ L^\#\end{array}\right] =  N
\left[\begin{array}{c} a \\  a^\#\end{array}\right],
\]
where $  N=
 \Delta(  N_1, N_2)$. In addition, we have an orthogonal {\em scattering matrix} $S$ which describes the interactions between the quantum fields. These quantities then lead to the following QSDEs which describe the dynamics of the quantum system under consideration:
\begin{eqnarray}
\label{qsde3} \left[\begin{array}{l} \dot a\\\dot a^\#\end{array}\right] &=&  F\left[\begin{array}{l} 
a\\ a^\#\end{array}\right] +  G
\left[\begin{array}{l} u
\\ u^{\#} \end{array}\right];  \nonumber \\
\left[\begin{array}{l} y
\\ y^\# \end{array}\right] &=&
 H \left[\begin{array}{l}  a\\
a^\#\end{array}\right] +  K \left[\begin{array}{l}
u
\\ u^{\#} \end{array}\right],\nonumber \\
\end{eqnarray}
where
\begin{eqnarray}
\label{FGHKform}  F =\Delta(  F_1,  F_2),
 &&
 G = \Delta( G_1,  G_2), \nonumber \\
 H = \Delta( H_1, H_2),
 &&  K = \Delta( K_1, K_2),
\end{eqnarray}
and
\begin{eqnarray}
\label{generalizedFGHK1}
 F &=& -\imath \Theta   M -\frac{1}{2} \Theta  N^\dagger J  N; ~~
 G = -\Theta   N^\dagger J\Delta(S,0); \nonumber \\
 H &=&  N;~~
 K = \Delta(S,0).
\end{eqnarray}

\noindent
{\bf Annihilation Operator Quantum Systems}
An important special case of the above class of linear quantum systems occurs when the QSDEs (\ref{qsde3}) can be described purely in terms of the vector of annihilation operators $a$; e.g., see \cite{MaP3,PET09A}. In this case, we consider 
Hamiltonian operators of the form
$
{\bf H} =
 a^\dagger  Ma 
$
 and coupling operator vectors of the form 
$
L = N a 
$
where $M$ is a Hermitian matrix and $N$ is a complex matrix. Also, we consider an orthogonal scattering matrix $S$.   In this case, we replace the commutation relations (\ref{CCR2}) by the commutation relations
\begin{eqnarray}
\label{CCR3} 
\left[a,a^\dagger\right]
&=& \Theta
\end{eqnarray}
where $\Theta$ is a positive-definite commutation matrix.
Then, the corresponding QSDEs are given by
\begin{eqnarray}
\label{qsde4}  \dot a &=&  F
a +  G u; ~~
y =
 H a +  K 
u
\end{eqnarray}
where
\begin{eqnarray}
\label{annihilationFGHK}
 F &=&  \Theta  \left( -\imath M +\frac{1}{2}   N^\dagger   N\right); ~~
 G = -\Theta   N^\dagger S; \nonumber \\
 H &=&  N; ~~
 K = S.
\end{eqnarray}

\section{Passive Linear Quantum Optical Networks}
Passive linear quantum optical networks consist of optical interconnections between the following passive optical components: optical cavities, beamsplitters, optical sources (lasers or vacuum sources), and optical sinks (detectors or unused optical outputs). We now describe each of these optical components in more details.

\begin{figure}[htbp]
\centering
\subfigure[Single mirror cavity]{\includegraphics[width=3cm]{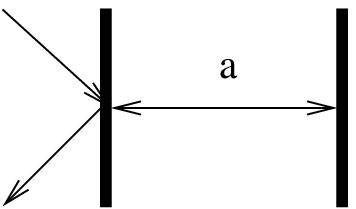}}\qquad
\subfigure[Two mirror cavity]{\includegraphics[width=3cm]{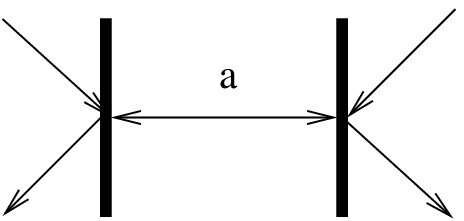}}\\
\subfigure[Two mirror butterfly cavity]{\includegraphics[width=3cm]{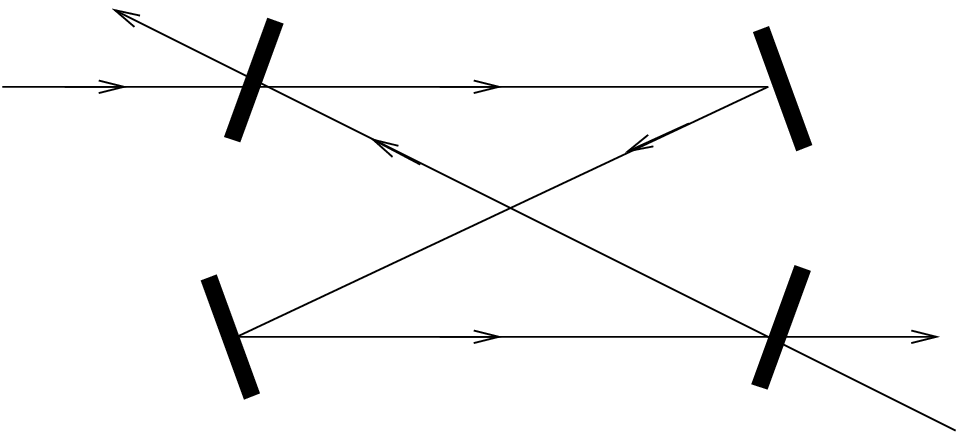}}\qquad
\subfigure[Three mirror ring cavity]{\includegraphics[width=3cm]{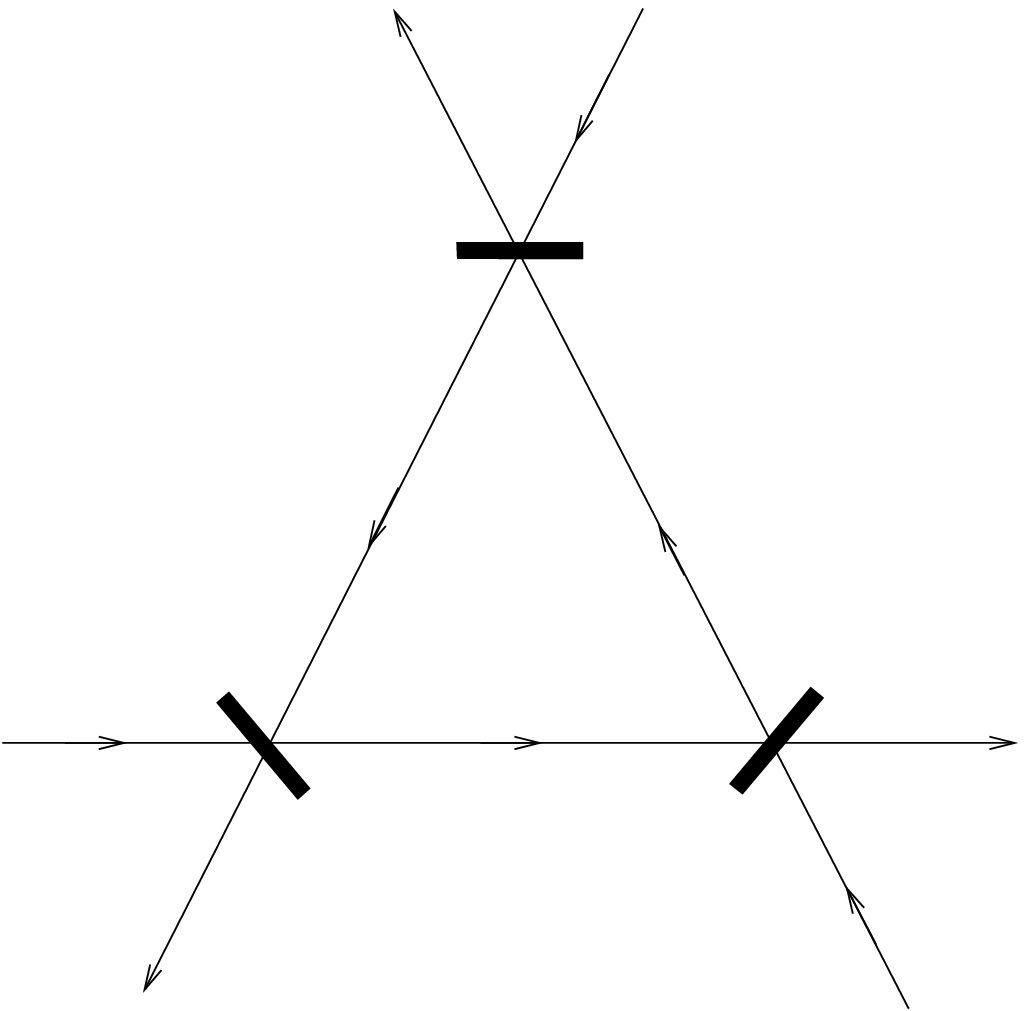}}
\caption{Some typical optical cavities.}%
\label{F1}%
\end{figure}

\noindent
{\bf Optical Cavities} \\
Optical cavities consist of a number of partially reflecting mirrors arranged in a suitable geometric configuration and coupled to a coherent light source such as a laser; e.g., see \cite{BR04,GZ00}. From the optical network point of view, we can categorize optical cavities according to the number of partially reflecting mirrors they contain.  Schematic diagrams for some typical optical cavities are shown  in Figure \ref{F1}.  Note that the single mirror cavity actually contains two mirrors but only one of the mirrors is partially reflecting. Similarly, the two mirror butterfly cavity actually contains four mirrors but only two of the mirrors are partially reflecting. In the sequel, we will ignore the fully reflecting mirrors in any cavity and only consider the partially reflecting mirrors. 

A
cavity with $m$ mirrors, can be described by
a QSDE of the form (\ref{qsde4}) as follows:
\begin{eqnarray}
\label{single_cavity}
\dot a &=& \left(-\frac{\gamma}{2} + \imath\Delta\right)a  - \sum_{i=1}^m \sqrt{\kappa_i} u_i;
  \nonumber \\
y_i &=& \sqrt{\kappa_i}a  + u_i,~~i=1,2,\ldots,m,
\end{eqnarray}
where 
\begin{equation}
\label{gamma}
\gamma =  \sum_{i=1}^m \kappa_i
\end{equation}
 and $a$ is an
annihilation operator associated with the cavity mode. The quantities
$\kappa_i \geq 0$, $i=1,2,\ldots,m$ are the {\em coupling
  coefficients}
which correspond to  the partially reflecting
mirrors which make up the cavity. Also,  $\Delta \in
\mathbb{R}$ corresponds to the {\em detuning} between the  cavity and the  coherent light source. 

\begin{figure}[htbp]
\centering
\psfrag{u1}{$u_1$}
\psfrag{u2}{$u_2$}
\psfrag{y1}{$y_1$}
\psfrag{y2}{$y_2$}
\includegraphics[width=3cm]{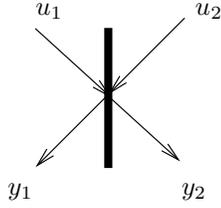}
\caption{Schematic diagram of a beamsplitter.}%
\label{F2}%
\end{figure}

\noindent
{\bf Beamsplitters} \\
A beamsplitter consists of a single partially reflective mirror as illustrated in Figure \ref{F2}. 
A beamsplitter is governed by the input-output relations
\begin{equation}
\label{beam1}
\left[\begin{array}{c}y_1\\y_2\end{array}\right]
= \left[\begin{array}{cc}\xi & -\sqrt{1-\xi^2} \\
-\sqrt{1-\xi^2} & - \xi\end{array}\right]
\left[\begin{array}{c}u_1\\u_2\end{array}\right]
\end{equation}
where $\xi \in (-1,1)$ is a parameter defining  the beamsplitter; e.g., see \cite{BR04}.
 In the sequel, it will be convenient to consider a beamsplitter as arising from a singular perturbation approximation applied to a two mirror cavity of the form shown in Figure \ref{F1}(b); see \cite{PET09A}. That is, we consider the following cavity equations of the form (\ref{single_cavity}):
\begin{eqnarray}
\label{2mirror1}
\dot a &=& \left(-\frac{ \pmb{\tilde \kappa} +\pmb{\bar \kappa}}{2} + \imath\Delta\right)a -  \sqrt{\pmb{\tilde \kappa}} u_1
-  \sqrt{\pmb{\bar \kappa}} u_2;
  \nonumber \\
y_1 &=& \sqrt{\pmb{\tilde \kappa}}a dt + u_1;~~
y_2 = \sqrt{\pmb{\bar \kappa}}a dt + u_2.
\end{eqnarray}
We now let
$
\tilde \kappa = \epsilon \pmb{\tilde \kappa},~~\bar \kappa = \epsilon \pmb{\bar \kappa},~~
\tilde a = \frac{a}{\sqrt \epsilon}
$
where $\epsilon  > 0$ is a given constant. Then, (\ref{2mirror1}) becomes
\begin{eqnarray}
\label{2mirror2}
\tilde a &=& \left(-\frac{ \tilde \kappa+ \bar \kappa}{2\epsilon} + \frac{\imath\Delta}{\epsilon}\right)\tilde a
 -  \frac{\sqrt{ \tilde \kappa}}{\epsilon} u_1
-  \frac{\sqrt{ \bar \kappa}}{\epsilon} u_2;
  \nonumber \\
y_1 &=& \sqrt{ \tilde \kappa}\tilde a  + u_1;~~
y_2 = \sqrt{ \bar \kappa}\tilde a  + u_2.
\end{eqnarray}
Letting $\epsilon \rightarrow 0$, we obtain
$
\frac{\tilde \kappa+ \bar \kappa - 2 \imath\Delta}{2} \tilde a = -\sqrt{ \tilde \kappa} u_1
-  \sqrt{ \bar \kappa} u_2
$
and hence
$
\tilde a = -\frac{2\sqrt{ \tilde \kappa}}{\tilde \kappa+ \bar \kappa - 2 \imath\Delta} u_1
- \frac{2\sqrt{ \bar \kappa}}{\tilde \kappa+ \bar \kappa - 2 \imath\Delta} u_2.
$
Substituting this into (\ref{2mirror2}) gives
\begin{eqnarray}
\label{beam2}
y_1 &=& \left(1- \frac{2\tilde \kappa}{\tilde \kappa+ \bar \kappa - 2 \imath\Delta} \right) u_1 
- \frac{2\sqrt{ \tilde \kappa \bar \kappa}}{\tilde \kappa+ \bar \kappa - 2 \imath\Delta} u_2;\nonumber \\
y_2 &=& -\frac{2\sqrt{\tilde \kappa\bar \kappa}}{\tilde \kappa+ \bar \kappa - 2 \imath\Delta} u_1
+\left(1- \frac{2\bar \kappa}{\tilde \kappa+ \bar \kappa - 2 \imath\Delta}\right) u_2.\nonumber \\
\end{eqnarray}
Letting, $\Delta = 0$, it follows that
\begin{equation}
\label{beam2a}
\left[\begin{array}{c}y_1\\y_2\end{array}\right]
= \left[\begin{array}{cc} \frac{\bar \kappa-\tilde \kappa}{\tilde \kappa+ \bar \kappa}& - \frac{2\sqrt{ \tilde \kappa \bar \kappa}}{\tilde \kappa+ \bar \kappa} \\
-\frac{2\sqrt{\tilde \kappa\bar \kappa}}{\tilde \kappa+ \bar \kappa} & \frac{\tilde \kappa-\bar \kappa}{\tilde \kappa+ \bar \kappa}\end{array}\right]
\left[\begin{array}{c}u_1\\u_2\end{array}\right].
\end{equation}
This equation is the same as (\ref{beam1}) when we let
\begin{equation}
\label{beam3}
\xi = \frac{\bar \kappa-\tilde \kappa}{\tilde \kappa+ \bar \kappa}.
\end{equation}

\noindent
{\bf Sources and Sinks} \\
Optical sources may be coherent sources such as a laser or a vacuum source which corresponds to no optical connection being made to a mirror input; e.g., see \cite{BR04}. We will represent both sources by the same schematic diagram as shown in Figure \ref{F2}(a).  Also, optical sinks may be detectors such as a homodyne detector (e.g., see \cite{BR04}) or they may  correspond to an unused optical output, which corresponds to no optical connection being made to a mirror output. We will represent both sinks by the same schematic diagram as shown in Figure \ref{F2}(b). Note that for the networks being considered, the number of sources will always be equal to the number of sinks. 

\begin{figure}[htbp]
\centering
\subfigure[Optical Source]{\includegraphics[width=3cm]{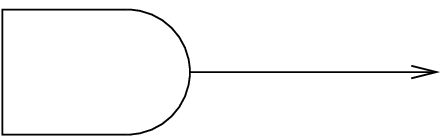}}\qquad
\subfigure[Optical Sink]{\includegraphics[width=3cm]{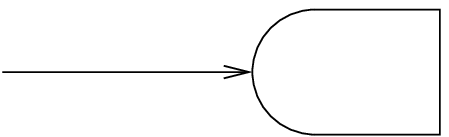}}
\caption{Schematic diagrams for optical sources and sinks.}%
\label{F3}%
\end{figure}

\begin{figure}[htbp]
\centering
\subfigure[Three mirror cavity decomposition]{\includegraphics[width=8cm]{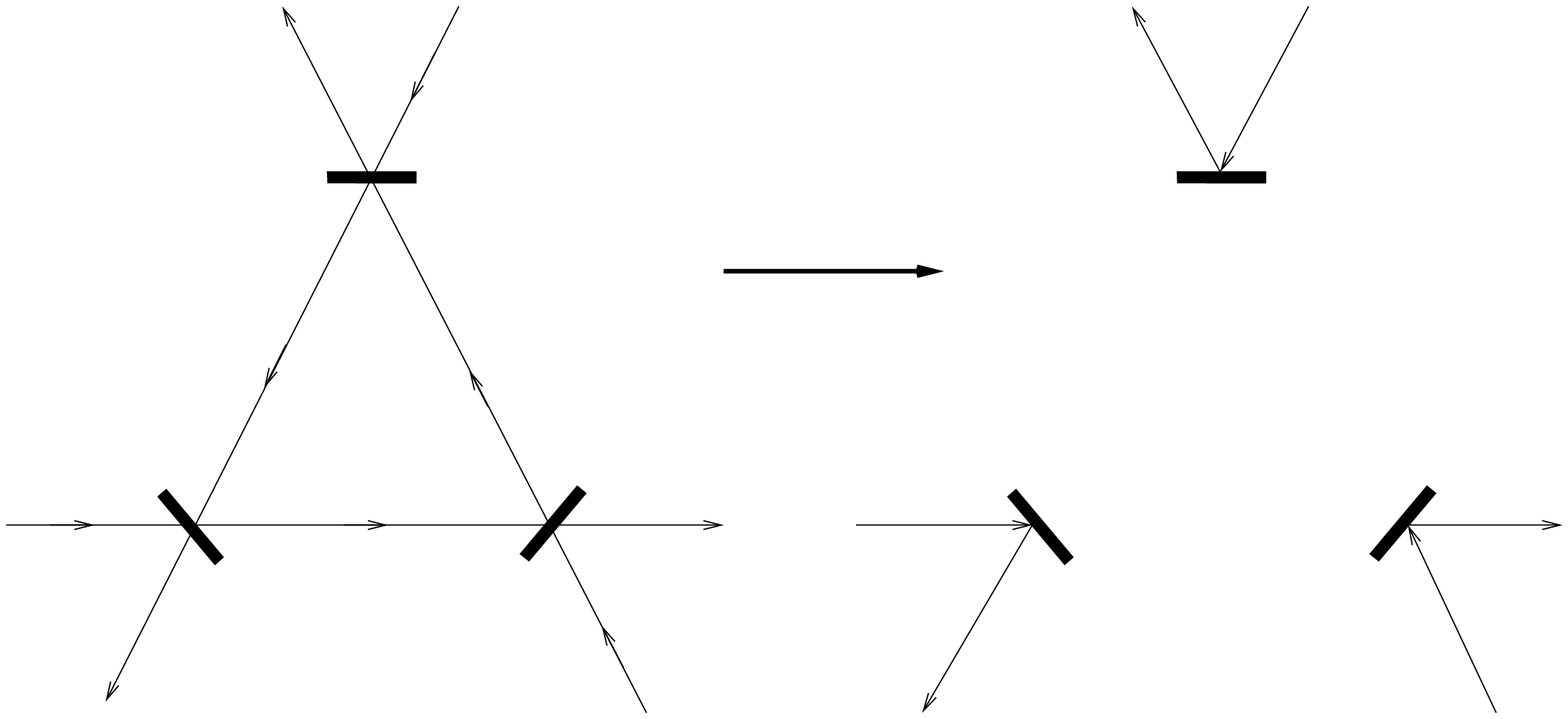}}\\
\subfigure[Beamsplitter decomposition]{\includegraphics[width=8cm]{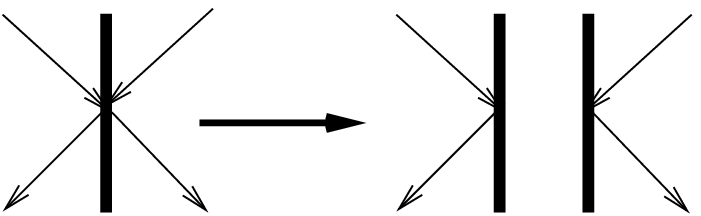}}
\caption{Decomposing a cavity and a beamsplitter into individual mirrors.}%
\label{F4}%
\end{figure}

\noindent
{\bf The Mirror Digraph} \\
We will consider the topology of an optical network to be represented by a directed graph referred to as the {\em mirror digraph}. To obtain the mirror digraph, each cavity in the network is decomposed into the mirrors that make up the cavity with an $m$-mirror cavity being decomposed into $m$ mirrors. Similarly, each beamsplitter is decomposed into two mirrors. This process is illustrated in Figure \ref{F4}. 

\begin{figure}[htbp]
\centering
\includegraphics[width=6cm]{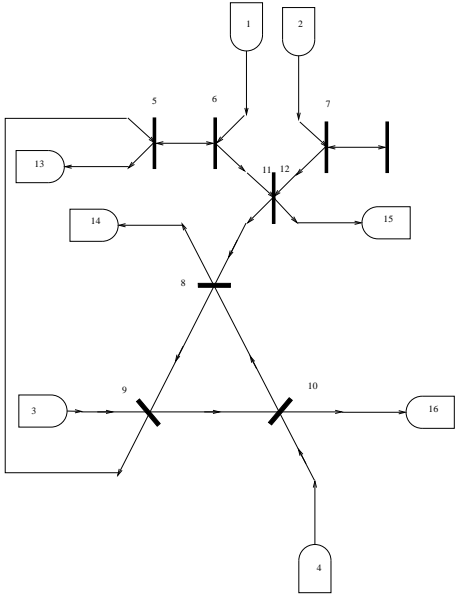}
\caption{A passive optical network.}%
\label{F5}%
\end{figure}

\begin{figure}[htbp]
\centering
\includegraphics[width=4cm]{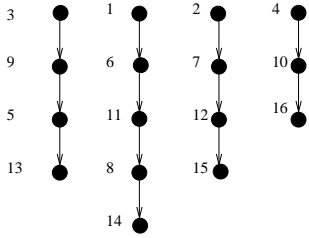}
\caption{Corresponding mirror digraph.}%
\label{F6}%
\end{figure}

Then a directed graph showing the interconnections of these mirrors, along with the optical sources and sinks is constructed. In this digraph, the nodes correspond to the mirrors or the optical sources and sinks. Also, the links in this graph correspond to the optical connections between the components. This process is illustrated in Figures \ref{F5} and \ref{F6} in which Figure \ref{F5} shows a passive optical network and Figure \ref{F6} shows the corresponding mirror digraph.

For a quantum optical network with $m$ sources, $n$ cavities including $n_m$ cavity mirrors, $k$ beamsplitters, and $m$ sinks, we will employ the following numbering convention. The sources will be numbered from 1 to $m$, the cavity mirrors will be numbered from $m+1$ to $m+n_m$, the beamsplitter mirrors will be numbered from $m+n_m+1$ to $m+n_m+2k$, and the sinks will be numbered from $m+n_m+2k+1$ to $2m+n_m+2k$. Associated with the mirror digraph is the corresponding {\em adjacency matrix} $A = \{a_{ij}\}$ defined so that $a_{ij} = 1$ is there is a link going from node $i$ to node $j$ and $a_{ij} = 0$ otherwise. Then, the adjacency matrix can be partitioned as follows corresponding to the different types of nodes:
\begin{equation}
\label{adj_part}
A = \left[\begin{array}{cccc}
A_{11} & A_{12} & A_{13} & A_{14}\\
A_{21} & A_{22} & A_{23} & A_{24}\\
A_{31} & A_{32} & A_{33} & A_{34}\\
A_{41} & A_{42} & A_{43} & A_{44}\\
\end{array}\right] \quad \left. \begin{array}{l} \mbox{sources,} \\ \mbox{cavities,} \\ \mbox{beamsplitters,} \\ \mbox{sinks.}
\end{array}\right. 
\end{equation}
Note that it follows from these definitions that the matrices $A_{11}$, $A_{21}$, $A_{41}$, $A_{42}$, $A_{43}$, and $A_{44}$ are all zero. 

The adjacency matrix corresponding to the passive optical network shown in Figure \ref{F5} and the corresponding mirror digraph shown in Figure \ref{F6} is 
\[
A = \left[\begin{array}{cccc|cccccc|cc|cccc}
0 &0& 0& 0& 0& 1& 0& 0& 0& 0& 0& 0& 0& 0& 0&0\\
0 &0& 0& 0& 0& 0& 1& 0& 0& 0& 0& 0& 0& 0& 0&0\\
0 &0& 0& 0& 0& 0& 0& 0& 1& 0& 0& 0& 0& 0& 0&0\\
0 &0& 0& 0& 0& 0& 0& 0& 0& 1& 0& 0& 0& 0& 0&0\\
\hline 
0 &0& 0& 0& 0& 0& 0& 0& 0& 0& 0& 0& 1& 0& 0&0\\
0 &0& 0& 0& 0& 0& 0& 0& 0& 0& 1& 0& 0& 0& 0&0\\
0 &0& 0& 0& 0& 0& 0& 0& 0& 0& 0& 1& 0& 0& 0&0\\
0 &0& 0& 0& 0& 0& 0& 0& 0& 0& 0& 0& 0& 1& 0&0\\
0 &0& 0& 0& 1& 0& 0& 0& 0& 0& 0& 0& 0& 0& 0&0\\
0 &0& 0& 0& 0& 0& 0& 0& 0& 0& 0& 0& 0& 0& 0&1\\
\hline 
0 &0& 0& 0& 0& 0& 0& 1& 0& 0& 0& 0& 0& 0& 0&0\\
0 &0& 0& 0& 0& 0& 0& 0& 0& 0& 0& 0& 0& 0& 1&0\\
\hline 
0 &0& 0& 0& 0& 0& 0& 0& 0& 0& 0& 0& 0& 0& 0&0\\
0 &0& 0& 0& 0& 0& 0& 0& 0& 0& 0& 0& 0& 0& 0&0\\
0 &0& 0& 0& 0& 0& 0& 0& 0& 0& 0& 0& 0& 0& 0&0\\
0 &0& 0& 0& 0& 0& 0& 0& 0& 0& 0& 0& 0& 0& 0&0
\end{array}\right].
\]
Hence in this example, $A_{13} = 0$, $A_{14} = 0$, $A_{33} = 0$,
\begin{eqnarray*}
A_{12} &=& \left[\begin{array}{cccccc}
0& 1& 0& 0& 0& 0\\
0& 0& 1& 0& 0& 0\\
0& 0& 0& 0& 1& 0\\
0& 0& 0& 0& 0& 1
\end{array}\right];~~
A_{22} = \left[\begin{array}{cccccc}
0& 0& 0& 0& 0& 0\\
0& 0& 0& 0& 0& 0\\
0& 0& 0& 0& 0& 0\\
0& 0& 0& 0& 0& 0\\
1& 0& 0& 0& 0& 0\\
0& 0& 0& 0& 0& 0
\end{array}\right];\\
A_{23} &=& \left[\begin{array}{cc}
0& 0\\
1& 0\\
0& 1\\
0& 0\\
0& 0\\
0& 0
\end{array}\right];~~
A_{24} = \left[\begin{array}{cccc}
1& 0& 0&0\\
0& 0& 0&0\\
0& 0& 0&0\\
0& 1& 0&0\\
0& 0& 0&0\\
0& 0& 0&1
\end{array}\right];\\
A_{32} &=& \left[\begin{array}{cccccc}
0& 0& 0& 1& 0& 0\\
0& 0& 0& 0& 0& 0
\end{array}\right];~~
A_{34} = \left[\begin{array}{cccc}
0& 0& 0&0\\
0& 0& 1&0\\
\end{array}\right].
\end{eqnarray*}

We will label the  field input  for the $i$th node of the mirror digraph as $u_i$ and the corresponding  field output  as $y_i$. In the case that the $i$th node of the mirror digraph corresponds to a source, there is no actual  field input but we will simply write $u_i=y_i$. Similarly, if the $i$th node of the mirror digraph corresponds to a sink, there is no actual  field output but we will simply write $y_i=u_i$. We then write
\[
\tilde u = \left[\begin{array}{c}
u_1 \\ \vdots \\u_{2m+n_m+2k}
\end{array}\right], \quad 
\tilde y = \left[\begin{array}{c}
y_1 \\ \vdots \\y_{2m+n_m+2k}
\end{array}\right].
\]
We now partition the vectors $\tilde u$ and $\tilde y$ according to the different types of nodes as follows:
\[
\tilde u = \left[\begin{array}{c}
\tilde u_1 \\
\tilde u_2 \\
\tilde u_3 \\
\tilde u_4
\end{array}\right]; \quad
\tilde y = \left[\begin{array}{c}
\tilde y_1 \\
\tilde y_2 \\
\tilde y_3 \\
\tilde y_4
\end{array}\right];
\quad \left. \begin{array}{l} \mbox{sources,} \\ \mbox{cavities,} \\ \mbox{beamsplitters,} \\ \mbox{sinks.}
\end{array}\right. 
\] 
Then using (\ref{adj_part}) and the definition of the adjacency matrix, we write
\begin{equation}
\label{adjacent}
\left[\begin{array}{c}
\tilde u_1 \\
\tilde u_2 \\
\tilde u_3 \\
\tilde u_4
\end{array}\right] = \left[\begin{array}{cccc}
I & 0 & 0 & 0\\
A_{12}^T & A_{22}^T & A_{32}^T & 0\\
A_{13}^T & A_{23}^T & A_{33}^T & 0\\
A_{14}^T & A_{24}^T & A_{34}^T & 0\\
\end{array}\right] 
\left[\begin{array}{c}
\tilde y_1 \\
\tilde y_2 \\
\tilde y_3 \\
\tilde y_4
\end{array}\right].
\end{equation}
Note that in writing these equations, we have ignored any phase shift which results from the light travelling from the output of node $i$ to the input of node $j$. We could allow for this phase shift by replacing the adjacency matrix (\ref{adj_part}) by a weighted adjacency matrix $A = \{a_{ij}\}$ in which any non-zero element is given by $a_{ij} = e^{\imath \theta_{ij}}$ where $\theta_{ij}$ is the phase shift in the light travelling from the output of node $i$ to the input of node $j$.

We will number the cavities from 1 to $n$. Then, the linear quantum optical network is also specified by a corresponding $n \times 2m+n_m+2k$ {\em cavity matrix} $C =  \{c_{ij}\}$ defined so that $c_{ij} = \sqrt{\kappa_j}$ if the mirror corresponding to the node $j$ in the mirror graph forms a part of cavity $i$. Here, $\kappa_j > 0$ is the coupling coefficient of the mirror corresponding to node $j$. It follows from this definition that the first $m$ and last $2k+m$ columns of the matrix $C$ will be zero since the corresponding nodes in the mirror graph do not correspond to mirrors in a cavity. Then, we can partition the matrix $C$ as follows corresponding to the different types of nodes:
\begin{eqnarray}
\label{C_partition}
&& \left.\begin{array}{cccc} 
\mbox{sources} & \mbox{cavities} & \mbox{beamsplitters} & \mbox{sinks}
\end{array}\right. \nonumber \\
C &=& \left[\begin{array}{cccc} 
0\quad\quad~~ & \tilde C \quad\quad~~ & 0 \quad\quad\quad\quad\quad & 0\quad\quad
\end{array}\right].
\end{eqnarray}

Also, it follows from this definition that we can write
\begin{equation}
\label{gamma_matrix}
CC^T = \tilde C\tilde C^T=\left[\begin{array}{cccc}
\gamma_1 & 0 & 0 \\
0 & \ddots & 0\\
0 & 0 & \gamma_n
\end{array}\right]
\end{equation}
where each quantity $\gamma_i$ is the sum of coupling coefficients of the mirrors forming cavity $i$ defined as in (\ref{gamma}).

In addition, we will define a diagonal $n \times n$   {\em detuning matrix} $D = \{d_{ij}\}$ defined so that $d_{ii} = \Delta_i$, the detuning of the $i$th cavity. The cavity matrix and the detuning matrix for the passive optical network shown in Figure \ref{F5} will be of the form
{\small
\begin{eqnarray*}
C &=& \left[\begin{array}{cccc|cccccc|cccccc}
0 &0& 0& 0& \sqrt\kappa_5& \sqrt\kappa_6& 0& 0& 0& 0& 0& 0& 0& 0& 0&0\\
0 &0& 0& 0& 0& 0& \sqrt\kappa_7& 0& 0& 0& 0& 0& 0& 0& 0&0\\
0 &0& 0& 0& 0& 0& 0& \sqrt\kappa_8& \sqrt\kappa_9& \sqrt\kappa_{10}& 0& 0& 0& 0& 0&0
\end{array}\right];\\
D &=& \left[\begin{array}{ccc}
\Delta_1 &0& 0 \\
0 &\Delta_2& 0 \\
0 &0& \Delta_3
\end{array}\right].
\end{eqnarray*}
}
Hence, 
\[
\tilde C = \left[\begin{array}{cccccc}
\sqrt\kappa_5& \sqrt\kappa_6& 0& 0& 0& 0\\
0& 0& \sqrt\kappa_7& 0& 0& 0\\
0& 0& 0& \sqrt\kappa_8& \sqrt\kappa_9& \sqrt\kappa_{10}
\end{array}\right].
\]

We will number the beamsplitters from 1 to $k$. Then, the linear quantum optical network is also specified by a corresponding $k \times 2m+n_m+2k$ {\em beamsplitter matrix} $B =  \{b_{ij}\}$ defined so that $b_{ij} = \sqrt \kappa_j$ if the mirror corresponding to the node $j$ in the mirror graph forms a part of beamsplitter $i$.  Here, $\kappa_j > 0$ is the coupling coefficient of the mirror corresponding to node $j$. In addition, we assume that each beamsplitter, which is represented by two mirrors in the mirror graph, is such that one mirror has a number $j \in \{m+n+1,\ldots, m+n+k\}$ and the other mirror has a number $j+k \in \{m+n+k+1,\ldots, m+n+2k\}$. It follows from this definition that the first $m+n$ and last $m$ columns of the matrix $B$ will be zero since the corresponding nodes in the mirror graph do not correspond to mirrors in a beamsplitter. Also, we can partition the matrix $B$ as follows corresponding to the different types of nodes:
\begin{eqnarray}
\label{B_partition}
&& \left.\begin{array}{ccccc} 
\mbox{sources} & \mbox{cavities} & \mbox{beamsplitters}& \mbox{beamsplitters} & \mbox{sinks}
\end{array}\right. \nonumber \\
B &=& \left[\begin{array}{ccccc} 
0\quad\quad~~ & 0 \quad\quad~~ & \tilde B \quad\quad\quad\quad\quad& \bar B \quad\quad\quad\quad\quad & 0
\end{array}\right].\nonumber \\
\end{eqnarray}
Hence, corresponding to each beamsplitter, there one non-zero entry in each of the square matrices $\tilde B$ and $\bar B$. For example, a network with three beamsplitters with parameters 
$\tilde \kappa_1$, $\bar \kappa_1$, $\tilde \kappa_2$, $\bar \kappa_2$, $\tilde \kappa_3$, $\bar \kappa_3$ respectively would have matrices
\[
\tilde B = \left[\begin{array}{ccc}\sqrt{\tilde \kappa_1} & 0 & 0 \\ 
0 & \sqrt{\tilde \kappa_2} & 0 \\
0 & 0 & \sqrt{\tilde \kappa_3} \end{array}\right]; ~~
\bar B = \left[\begin{array}{ccc}\sqrt{\bar \kappa_1} & 0 & 0 \\ 
0 & \sqrt{\bar \kappa_2} & 0 \\
0 & 0 & \sqrt{\bar \kappa_3} \end{array}\right].
\]

The coupling coefficients in the beamsplitter matrix form the parameters $\xi$ in the corresponding beamsplitter equations of the form (\ref{beam1}) according to the formula (\ref{beam3}); i.e., 
\[
\xi_i = \frac{\kappa_{\tilde j_i}-\kappa_{\bar j_i}}{\kappa_{\tilde j_i}+ \kappa_{\bar j_i}}
\]
where the mirrors corresponding to the nodes ${\tilde j_i}$ and ${\bar j_i}$ in the mirror digraph make up the $i$th beamsplitter. Also, it follows from the definition of $B$ that we can write
\begin{equation}
\label{BBt_matrix}
BB^T = \tilde B\tilde B^T+\bar B\bar B^T=\left[\begin{array}{cccc}
\kappa_{\tilde j_1}+ \kappa_{\bar j_1} & 0 & 0 \\
0 & \ddots & 0\\
0 & 0 & \kappa_{\tilde j_k}+ \kappa_{\bar j_k}
\end{array}\right].
\end{equation}

Note that there is some redundancy in the choice of the parameters $\kappa_{\tilde j_i} > 0$ and $\kappa_{\bar j_i} > 0$ for a given beamsplitter since its behaviour is defined by a single parameter $\xi_i \in (-1,1)$. The beamsplitter matrix  for the passive optical network shown in Figure \ref{F5} will be of the form
\[
B = \left[\begin{array}{cccccccccc|cc|cccc}
0 &0& 0& 0& 0& 0& 0& 0& 0& 0& \sqrt{\kappa_{11}}& \sqrt{\kappa_{12}}& 0& 0& 0&0
\end{array}\right].
\]
Hence, 
\[
\tilde B = \sqrt{\kappa_{11}}; \quad \bar B = \sqrt{\kappa_{12}}.
\]

\begin{figure}[htbp]
\centering
\includegraphics[width=8cm]{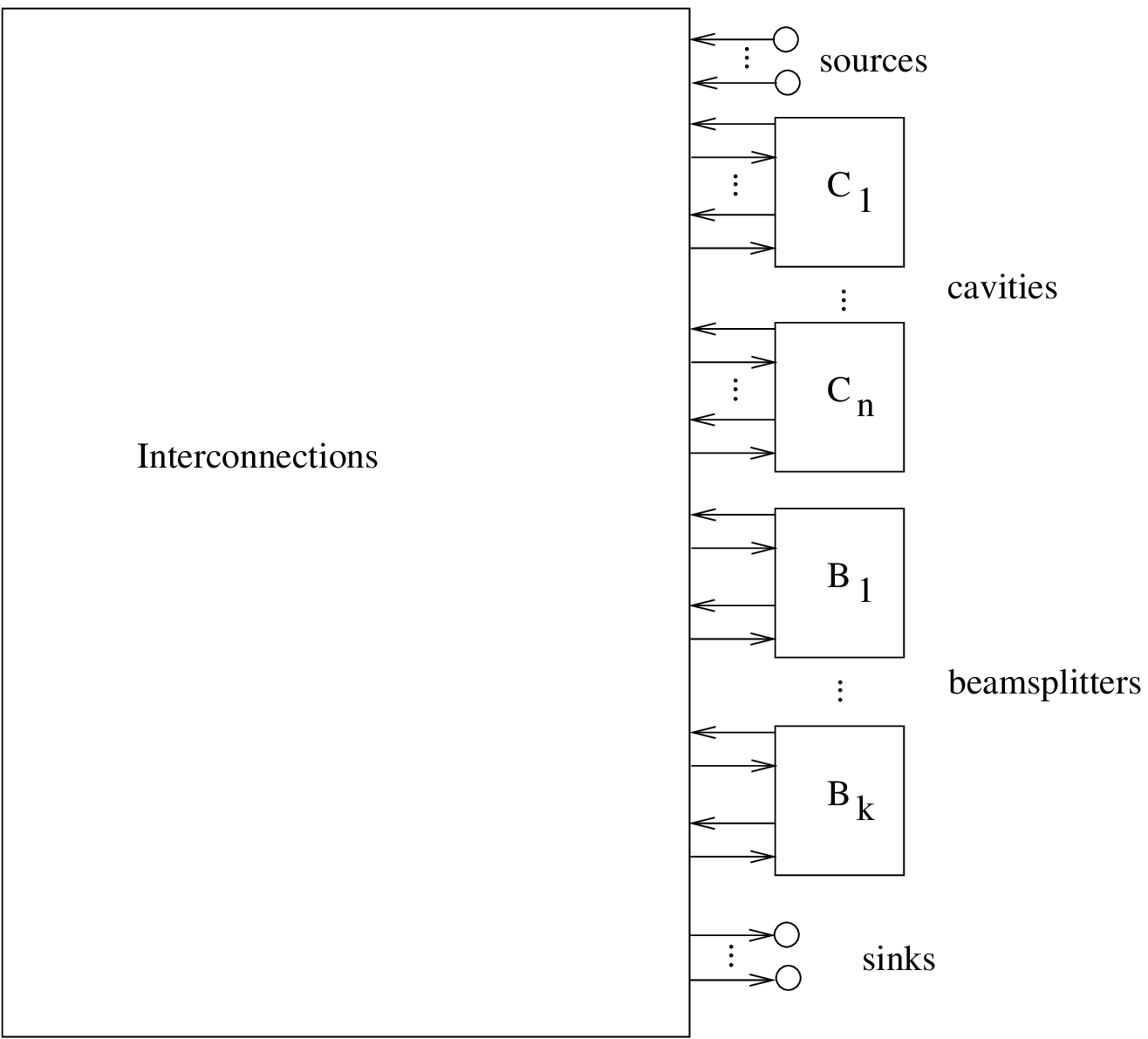}
\caption{Component extraction for a quantum optical network.}%
\label{F7}%
\end{figure}

\noindent
{\bf Writing the QSDEs for a passive quantum optical network} \\
Together the matrices $A$, $C$, $D$, $B$ completely specify the a given passive quantum optical network. We will now derive QSDEs of the form (\ref{qsde3}) in terms of these matrices to describe a given network. To do this, we first extract all of the sources, sinks, cavities and beamsplitters from the network in a similar fashion to the reactance extraction process which is carried out in circuit theory analysis; e.g., see \cite{AV73}. This is illustrated in Figure \ref{F7}. 

In this picture, an $\ell$ mirror cavity is regarded as an $\ell$ port network with $\ell$ inputs and $\ell$ outputs using a scattering framework; e.g., see \cite{AV73}. Also, a beamsplitter is regarded as a two port network. 

{\bf Cavity Equations} We now consider the QSDEs (\ref{qsde3}) corresponding to the $i$th cavity. Letting $a_i$ be the annihilation operator corresponding to the $i$th cavity, it follows from (\ref{single_cavity}) and the definitions of the matrices $C$ and $D$ that we can write
\begin{eqnarray}
\label{cavity_i}
\dot a_i &=& \left(-\frac{\gamma_i}{2} + \imath d_{ii}\right)a_i  - \sum_{j=1}^{2m+n_m+2k} c_{ij} u_j;\nonumber \\
y_j &=& c_{ji}a_i + u_j 
\end{eqnarray}
for $j \in \{1,\ldots, 2m+n_m+2k\}$  such that  $c_{ji} \neq 0$.
Then, we define the vector of system variables
\[
a = \left[\begin{array}{c}
a_1 \\
\vdots \\
a_n
\end{array}\right]
\]
and use (\ref{gamma_matrix}) to write all of the equations (\ref{cavity_i}) in matrix form as follows:
\begin{eqnarray}
\label{cavity_equations}
\dot a &=& \left(-\frac{1}{2}\tilde C\tilde C^T + \imath D\right)a - \tilde C \tilde u_2;\nonumber \\
\tilde y_2 &=& \tilde C^Ta + \tilde u_2. 
\end{eqnarray}

{\bf Beamsplitter Equations} We now consider the relationship between the inputs to the beamsplitters $\tilde u_3$ and the outputs of the beamsplitters $\tilde y_3$. We first consider a single beamsplitter with parameters $\tilde \kappa$ and $\bar \kappa$. 
That is, we let $\tilde B = \sqrt{ \tilde \kappa}$ 
and $\bar B = \sqrt{ \bar \kappa}$. Then, using (\ref{BBt_matrix})  we calculate
 \begin{eqnarray*}
 &&-\left[\begin{array}{cc} \tilde B & \bar B \end{array}\right]^T
 \left( \tilde B\tilde B^T+\bar B\bar B^T\right)^{-1} \left[\begin{array}{cc} \tilde B & \bar B \end{array}\right] \\
&&+
 \left[\begin{array}{cc} -\bar B & \tilde B \end{array}\right]^T
 \left( \tilde B\tilde B^T+\bar B\bar B^T\right)^{-1} \left[\begin{array}{cc} -\bar B & \tilde B \end{array}\right]\\
  &=& \left[\begin{array}{cc} \frac{\bar \kappa-\tilde \kappa}{\tilde \kappa+ \bar \kappa}& - \frac{2\sqrt{ \tilde \kappa \bar \kappa}}{\tilde \kappa+ \bar \kappa} \\
 -\frac{2\sqrt{\tilde \kappa\bar \kappa}}{\tilde \kappa+ \bar \kappa} & \frac{\tilde \kappa-\bar \kappa}{\tilde \kappa+ \bar \kappa}\end{array}\right]
 \end{eqnarray*}
  which is the same as the matrix in (\ref{beam2a}). We now extend this formula to the case of $k$ beamsplitters and obtain
\begin{equation}
\label{beam4}
\tilde y_3 = \hat B \tilde u_3
\end{equation}
where
\begin{eqnarray*}
\hat B &=& -\left[\begin{array}{cc} \tilde B & \bar B \end{array}\right]^T
 \left( \tilde B\tilde B^T+\bar B\bar B^T\right)^{-1} \left[\begin{array}{cc} \tilde B & \bar B \end{array}\right] \\
&&+
 \left[\begin{array}{cc} -\bar B & \tilde B \end{array}\right]^T
 \left( \tilde B\tilde B^T+\bar B\bar B^T\right)^{-1} \left[\begin{array}{cc} -\bar B & \tilde B \end{array}\right].
\end{eqnarray*}
For the passive optical network shown in Figure \ref{F5}, we calculate
\[
\hat B = \left[\begin{array}{cc} \frac{\kappa_{12}-\kappa_{11}}{\kappa_{11}+ \kappa_{12}}& - \frac{2\sqrt{ \kappa_{11} \kappa_{12}}}{\kappa_{11}+ \kappa_{12}} \\
 -\frac{2\sqrt{\kappa_{11}\kappa_{12}}}{\kappa_{11}+ \kappa_{12}} & \frac{\kappa_{11}-\kappa_{12}}{\kappa_{11}+ \kappa_{12}}\end{array}\right].
\]

We now combine the equations (\ref{adjacent}), (\ref{cavity_equations}), (\ref{beam4}) to obtain a set of QSDEs of the form (\ref{qsde4}) which describes the complete network. In order to do this, we require that the network satisfies the following assumption:

\begin{assumption}
\label{A1}
The matrix $I - \left[\begin{array}{cc}A_{22}^T & A_{32}^T \hat B\\ A_{23}^T & A_{33}^T\hat B
\end{array}\right]$ is nonsingular. 
\end{assumption}
This assumption will be satisfied if the network does not contain any algebraic loops. If this assumption is not satisfied, the network will need to be modelled by a set of stochastic algebraic-differential equations. 

It follows from (\ref{adjacent}) that we can write
\begin{eqnarray*}
\tilde u_2 &=& A_{12}^T\tilde u_1 + A_{22}^T\tilde y_2 + A_{32}^T\tilde y_3; \\
\tilde u_3 &=& A_{13}^T\tilde u_1 + A_{23}^T\tilde y_2 + A_{33}^T\tilde y_3.
\end{eqnarray*}
Combining this with (\ref{beam4}) and the second equation in (\ref{cavity_equations}), we obtain:
\[
\left[\begin{array}{cc}\tilde u_2\\ \tilde u_3\end{array}\right]
= \left[\begin{array}{c}A_{12}^T\\ A_{13}^T\end{array}\right]\tilde u_1
+ \left[\begin{array}{c}A_{22}^T\\ A_{23}^T\end{array}\right]\tilde C^T a 
+\left[\begin{array}{cc}A_{22}^T & A_{32}^T \hat B\\ A_{23}^T & A_{33}^T\hat B
\end{array}\right]\left[\begin{array}{cc}\tilde u_2\\ \tilde u_3\end{array}\right].
\]
Now using Assumption \ref{A1}, it follows that we can write
\begin{eqnarray*}
\left[\begin{array}{cc}\tilde u_2\\ \tilde u_3\end{array}\right]
&=& \left(I - \left[\begin{array}{cc}A_{22}^T & A_{32}^T \hat B\\ A_{23}^T & A_{33}^T\hat B
\end{array}\right]\right)^{-1}
\left[\begin{array}{c}A_{12}^T\\ A_{13}^T\end{array}\right]\tilde u_1\\
&&+\left(I - \left[\begin{array}{cc}A_{22}^T & A_{32}^T \hat B\\ A_{23}^T & A_{33}^T\hat B
\end{array}\right]\right)^{-1}
\left[\begin{array}{c}A_{22}^T\\ A_{23}^T\end{array}\right]\tilde C^T a .
\end{eqnarray*}
Substituting this into (\ref{cavity_equations}) and using the last equation in (\ref{adjacent}), we obtain the following QSDEs of the form (\ref{qsde4}) which describe the network:
\begin{eqnarray*}
\dot a &=& \left(-\frac{1}{2}\tilde C\tilde C^T + \imath D\right)a \\
&&- \left[\begin{array}{cc}\tilde C & 0 \end{array}\right]\left(I - \left[\begin{array}{cc}A_{22}^T & A_{32}^T \hat B\\ A_{23}^T & A_{33}^T\hat B
\end{array}\right]\right)^{-1}
\left[\begin{array}{c}A_{22}^T\\ A_{23}^T\end{array}\right]\tilde C^T a\\
&&- \left[\begin{array}{cc}\tilde C & 0 \end{array}\right] \left(I - \left[\begin{array}{cc}A_{22}^T & A_{32}^T \hat B\\ A_{23}^T & A_{33}^T\hat B
\end{array}\right]\right)^{-1}
\left[\begin{array}{c}A_{12}^T\\ A_{13}^T\end{array}\right]\tilde u_1;\\
\tilde y_4 &=& A_{24}^T\tilde C^Ta \\
&&+\left[\begin{array}{cc}A_{24}^T & A_{34}^T\hat B\end{array}\right]\left(I - \left[\begin{array}{cc}A_{22}^T & A_{32}^T \hat B\\ A_{23}^T & A_{33}^T\hat B
\end{array}\right]\right)^{-1}
\left[\begin{array}{c}A_{22}^T\\ A_{23}^T\end{array}\right]\tilde C^T a\\
&& + \left[\begin{array}{cc}A_{24}^T & A_{34}^T\hat B\end{array}\right]\left(I - \left[\begin{array}{cc}A_{22}^T & A_{32}^T \hat B\\ A_{23}^T & A_{33}^T\hat B
\end{array}\right]\right)^{-1}
\left[\begin{array}{c}A_{12}^T\\ A_{13}^T\end{array}\right]\tilde u_1\\
&&+ A_{14}^T\tilde u_1.
\end{eqnarray*}

From this, we can also use the formulas (4) and (5) in \cite{PET09A} with $\Theta = I$ to calculate the corresponding matrices $S$, $N$, $M$ in the $(S,L,H)$ description of this system. This yields
{\small
\begin{eqnarray*}
S &=& \left[\begin{array}{cc}A_{24}^T & A_{34}^T\hat B\end{array}\right]\left(I - \left[\begin{array}{cc}A_{22}^T & A_{32}^T \hat B\\ A_{23}^T & A_{33}^T\hat B
\end{array}\right]\right)^{-1}
\left[\begin{array}{c}A_{12}^T\\ A_{13}^T\end{array}\right]+ A_{14}^T;\\
N &=&\left[\begin{array}{l}A_{24}^T+\\
\left[\begin{array}{cc}A_{24}^T & A_{34}^T\hat B\end{array}\right]\left(I - \left[\begin{array}{cc}A_{22}^T & A_{32}^T \hat B\\ A_{23}^T & A_{33}^T\hat B
\end{array}\right]\right)^{-1}
\left[\begin{array}{c}A_{22}^T\\ A_{23}^T\end{array}\right]\end{array}\right]\tilde C^T;\\
M &=& -D \\
&&-\frac{\imath}{2} \tilde C \left[\begin{array}{l}\left[\begin{array}{cc}I & 0\end{array}\right]
\left(I - \left[\begin{array}{cc}A_{22}^T & A_{32}^T \hat B\\ A_{23}^T & A_{33}^T\hat B
\end{array}\right]\right)^{-1}
\left[\begin{array}{c}A_{22}^T\\ A_{23}^T\end{array}\right] -\\
\left[\begin{array}{cc}A_{22} & A_{23}\end{array}\right]
\left(I - \left[\begin{array}{cc}A_{22} & A_{23} \\ \hat B^TA_{32} & \hat B^T A_{33}
\end{array}\right]\right)^{-1}
\left[\begin{array}{c}I\\ 0\end{array}\right]
\end{array} \right]\tilde C^T. 
\end{eqnarray*}
}
For the passive optical network shown in Figure \ref{F5} we calculate the matrices $F$, $G$, $H$, $K$ in the corresponding QSDEs of the form (\ref{qsde4}) to be
\begin{eqnarray*}
F&=& \left[\begin{array}{ll}
\imath \Delta_1 - \frac{\kappa_6}{2} - \frac{\kappa_5}{2}&0\\
0&\imath \Delta_2 - \frac{\kappa_7}{2}\\
 \frac{\sqrt{\kappa_6}\sqrt{\kappa_8} \left(\kappa_{11} - \kappa_{12}\right)}{\kappa_{11} + \kappa_{12}}& 
 \frac{2 \sqrt{\kappa_7} \sqrt{\kappa_8}\sqrt{\kappa_{11}\kappa_{12}}}{\kappa_{11} + \kappa_{12}}
\end{array}\right.\\
&& \hspace{4cm}\left.\begin{array}{r}
-\sqrt{\kappa_5} \sqrt{\kappa_9}\\
0\\
 \imath \Delta_3 - \frac{\kappa_8}{2} - \frac{\kappa_9}{2} - \frac{\kappa_{10}}{2}
\end{array}\right];\\
G&=& \left[\begin{array}{rrrr}
-\sqrt{\kappa_6}&0& -\sqrt{\kappa_5}&0\\
0&-\sqrt{\kappa_7}&0&0\\
\frac{\sqrt{\kappa_8}\left(\kappa_{11} - \kappa_{12}\right)}{\kappa_{11} + \kappa_{12}}& 
\frac{2\sqrt{\kappa_8}\sqrt{\kappa_{11}\kappa_{12}}}{\kappa_{11} + \kappa_{12}}&
 -\sqrt{\kappa_9}& -\sqrt{\kappa_{10}}
\end{array}\right];\\
H&=&\left[\begin{array}{rrr}
\sqrt{\kappa_5}&0&\sqrt{\kappa_9}\\
 -\frac{\sqrt{\kappa_6}\left(\kappa_{11} - \kappa_{12}\right)}{\kappa_{11} + \kappa_{12}}&
 -\frac{2\sqrt{\kappa_7}\sqrt{\kappa_{11}\kappa_{12}}}{\kappa_{11} + \kappa_{12}}&  \sqrt{\kappa_8}\\
  -\frac{2\sqrt{\kappa_6}\sqrt{\kappa_{11}\kappa_{12}}}{\kappa_{11} + \kappa_{12}}& 
  \frac{\sqrt{\kappa_7}\left(\kappa_{11} - \kappa_{12}\right)}{\kappa_{11} + \kappa_{12}}&0\\
 0&0& \sqrt{\kappa_{10}}
\end{array}\right];\\
K&=& \left[\begin{array}{rrrr}
0&0& 1& 0\\
-\frac{\kappa_{11} - \kappa_{12}}{\kappa_{11} + \kappa_{12}}& 
-\frac{2\sqrt{\kappa_{11}\kappa_{12}}}{\kappa_{11} + \kappa_{12}}& 0& 0\\
-\frac{2\sqrt{\kappa_{11}\kappa_{12}}}{\kappa_{11} + \kappa_{12}}&
\frac{\kappa_{11} - \kappa_{12}}{\kappa_{11} + \kappa_{12}}& 0& 0\\
0&0& 0& 1
\end{array}\right].
\end{eqnarray*}

\noindent
{\bf  QSDEs for cavity only networks} \\
In this special case, we replace Assumption \ref{A1} by the following assumption.
\begin{assumption}
\label{A2}
The matrix $I - A_{22}^T$ is nonsingular. 
\end{assumption}
This assumption will be satisfied provided that none of the cavity mirrors have their output directly connected to their input. 

In this case, it follows from (\ref{adjacent}) that we can write
\begin{eqnarray*}
\tilde u_2 &=& A_{12}^T\tilde u_1 + A_{22}^T\tilde y_2.
\end{eqnarray*}
Combining this with  the second equation in (\ref{cavity_equations}), we obtain:
\[
\tilde u_2
= A_{12}^T\tilde u_1
+ A_{22}^T\tilde C^T a 
+A_{22}^T\tilde u_2.
\]
Now using Assumption \ref{A2}, it follows that we can write
\begin{eqnarray*}
\tilde u_2
&=& \left(I - A_{22}^T\right)^{-1}
A_{12}^T\tilde u_1
+\left(I - A_{22}^T\right)^{-1}
A_{22}^T\tilde C^T a .
\end{eqnarray*}
Substituting this into (\ref{cavity_equations}) and using the last equation in (\ref{adjacent}), we obtain the following QSDEs of the form (\ref{qsde4}) which describes the network:
\begin{eqnarray*}
\dot a &=& \left(-\frac{1}{2}\tilde C\tilde C^T + \imath D - \tilde C\left(I -A_{22}^T\right)^{-1}
A_{22}^T\tilde C^T\right)a\\
&&- \tilde C \left(I - A_{22}^T \right)^{-1}
A_{12}^T\tilde u_1;\\
\tilde y_4 &=& A_{24}^T\left(I - A_{22}^T\right)^{-1}\tilde C^T a \\&&
+ \left(A_{24}^T\left(I - A_{22}^T \right)^{-1}
A_{12}^T
+ A_{14}^T\right)\tilde u_1.
\end{eqnarray*}
From this, we can also use the formulas (4) and (5) in \cite{PET09A} with $\Theta = I$ to calculate the corresponding matrices $S$, $N$, $M$ in the $(S,L,H)$ description of this system. This yields
\begin{eqnarray*}
S &=& A_{24}^T\left(I - A_{22}^T \right)^{-1}
A_{12}^T+ A_{14}^T;\\
N &=& A_{24}^T\left(I - A_{22}^T\right)^{-1}\tilde C^T;\\
M &=& -D +\frac{\imath}{2} \tilde C \left(I -A_{22}^T\right)^{-1}\left(A_{22}-A_{22}^T\right)\left(I -A_{22}\right)^{-1}\tilde C^T. 
\end{eqnarray*}
\noindent
{\bf  Equations for beamsplitter only networks} \\
In this special case, we replace Assumption \ref{A1} by the following assumption.
\begin{assumption}
\label{A3}
The matrix $I - A_{33}^T\hat B$ is nonsingular. 
\end{assumption}
This assumption will be satisfied if the beamsplitter network does not contain any algebraic loops. 

It follows from (\ref{adjacent}) that we can write
\begin{eqnarray*}
\tilde u_3 &=& A_{13}^T\tilde u_1  + A_{33}^T\tilde y_3.
\end{eqnarray*}
Combining this with (\ref{beam4})  we obtain:
\[
\tilde u_3
= A_{13}^T\tilde u_1
+A_{33}^T\hat B
 \tilde u_3.
\]
Now using Assumption \ref{A3}, it follows that we can write
\begin{eqnarray*}
 \tilde u_3
&=& \left(I - A_{33}^T\hat B
\right)^{-1}
A_{13}^T\tilde u_1.
\end{eqnarray*}
Substituting this into  the last equation in (\ref{adjacent}), we obtain the following equation which describes the network:
\begin{eqnarray*}
\tilde y_4 &=&  A_{34}^T\hat B\left(I - A_{33}^T\hat B\right)^{-1}
A_{13}^T\tilde u_1+ A_{14}^T\tilde u_1.
\end{eqnarray*}

From this, we can also calculate the corresponding parameters in the $(S,L,H)$ description of this system to be
\begin{eqnarray*}
S &=& A_{34}^T\hat B\left(I - A_{33}^T\hat B\right)^{-1}
A_{13}^T+ A_{14}^T;\\
L &=& 0;\\
{\bf H} &=& 0.
\end{eqnarray*}
\section{Active Linear Quantum Optical Networks}
To consider active linear quantum networks, we extend the analysis given in the previous section to allow for dynamic squeezers (degenerate parametric amplifiers) instead of passive cavities; e.g., see \cite{BR04,GZ00,PET10Ca}. Indeed, by including a nonlinear optical element inside an $m$ mirror cavity, an $m$ mirror optical squeezer can be obtained. By using suitable linearizations and approximations, such an optical squeezer can be
 described by
a complex quantum stochastic differential equation as follows:
\begin{eqnarray}
\label{single_cavity}
da &=& \left(-\frac{\gamma}{2} + i\Delta\right)a dt -\chi a^* dt - \sum_{i=1}^m \sqrt{\kappa_i} du_i;
  \nonumber \\
dy_i &=& \sqrt{\kappa_i}a dt + du_i,~~i=1,2,\ldots,m,
\end{eqnarray}
where $\kappa_i > 0$ for $i=1,2,\ldots,m$, $\gamma =  \sum_{i=1}^m \kappa_i$, $\chi \in \mathbb{C}$, $\Delta \in
\mathbb{R}$, and $a$ is a single
annihilation operator associated with the cavity mode; e.g., see \cite{BR04,GZ00,PET10Ca}. Also, the quantity $\chi$ corresponds to the level of squeezing achieved in the squeezer. Note that in the case that $\chi = 0$, these equations reduce to those of a passive $m$ mirror cavity and hence, without loss of generality, we can assume that all cavities in the network are of this form. Hence, we can proceed in exactly the same fashion as in the previous section with the only addition being that we define an $n \times n$   diagonal  matrix $\mathcal{X} = \{x_{ij}\}$ defined so that $x_{ii} = \chi_i$, the squeezing parameter of the $i$th cavity. Then, the equations corresponding to equations (\ref{cavity_equations}), which describe the dynamics of all of the squeezers in the network, become
\begin{eqnarray}
\label{squeezer_equations}
\dot a &=& \left(-\frac{1}{2}\tilde C\tilde C^T + \imath D\right)a -\mathcal{X}a^\# - \tilde C \tilde u_2;\nonumber \\
\dot a^\# &=& -\mathcal{X}^\#a +\left(-\frac{1}{2}\tilde C\tilde C^T - \imath D\right)a^\#  - \tilde C \tilde u_2^\#;\nonumber \\
\tilde y_2 &=& \tilde C^Ta + \tilde u_2;\nonumber \\
\tilde y_2^\# &=& \tilde C^Ta^\# + \tilde u_2^\#. 
\end{eqnarray}
Assuming that Assumption \ref{A1} is satisfied, these equations are then combined with equations (\ref{adjacent}) and (\ref{beam4}) to obtain the following QSDEs of the form (\ref{qsde3}) which describe the complete active linear quantum optical network:
\begin{eqnarray*}
\dot a &=& \left(-\frac{1}{2}\tilde C\tilde C^T + \imath D\right)a \\
&&- \left[\begin{array}{cc}\tilde C & 0 \end{array}\right]\left(I - \left[\begin{array}{cc}A_{22}^T & A_{32}^T \hat B\\ A_{23}^T & A_{33}^T\hat B
\end{array}\right]\right)^{-1}
\left[\begin{array}{c}A_{22}^T\\ A_{23}^T\end{array}\right]\tilde C^T a\\
&&-\mathcal{X}a^\#\\
&&- \left[\begin{array}{cc}\tilde C & 0 \end{array}\right] \left(I - \left[\begin{array}{cc}A_{22}^T & A_{32}^T \hat B\\ A_{23}^T & A_{33}^T\hat B
\end{array}\right]\right)^{-1}
\left[\begin{array}{c}A_{12}^T\\ A_{13}^T\end{array}\right]\tilde u_1;\\
\dot a^\# &=& -\mathcal{X}^\#a+\left(-\frac{1}{2}\tilde C\tilde C^T - \imath D\right)a^\# \\
&&- \left[\begin{array}{cc}\tilde C & 0 \end{array}\right]\left(I - \left[\begin{array}{cc}A_{22}^T & A_{32}^T \hat B\\ A_{23}^T & A_{33}^T\hat B
\end{array}\right]\right)^{-1}
\left[\begin{array}{c}A_{22}^T\\ A_{23}^T\end{array}\right]\tilde C^T a^\#\\
&&- \left[\begin{array}{cc}\tilde C & 0 \end{array}\right] \left(I - \left[\begin{array}{cc}A_{22}^T & A_{32}^T \hat B\\ A_{23}^T & A_{33}^T\hat B
\end{array}\right]\right)^{-1}
\left[\begin{array}{c}A_{12}^T\\ A_{13}^T\end{array}\right]\tilde u_1^\#;\\
\tilde y_4 &=& A_{24}^T\tilde C^Ta \\
&&+\left[\begin{array}{cc}A_{24}^T & A_{34}^T\hat B\end{array}\right]\left(I - \left[\begin{array}{cc}A_{22}^T & A_{32}^T \hat B\\ A_{23}^T & A_{33}^T\hat B
\end{array}\right]\right)^{-1}
\left[\begin{array}{c}A_{22}^T\\ A_{23}^T\end{array}\right]\tilde C^T a\\
&& + \left[\begin{array}{cc}A_{24}^T & A_{34}^T\hat B\end{array}\right]\left(I - \left[\begin{array}{cc}A_{22}^T & A_{32}^T \hat B\\ A_{23}^T & A_{33}^T\hat B
\end{array}\right]\right)^{-1}
\left[\begin{array}{c}A_{12}^T\\ A_{13}^T\end{array}\right]\tilde u_1\\
&&+ A_{14}^T\tilde u_1;\\
\tilde y_4^\# &=& A_{24}^T\tilde C^Ta^\# \\
&&+\left[\begin{array}{cc}A_{24}^T & A_{34}^T\hat B\end{array}\right]\left(I - \left[\begin{array}{cc}A_{22}^T & A_{32}^T \hat B\\ A_{23}^T & A_{33}^T\hat B
\end{array}\right]\right)^{-1}
\left[\begin{array}{c}A_{22}^T\\ A_{23}^T\end{array}\right]\tilde C^T a^\#\\
&& + \left[\begin{array}{cc}A_{24}^T & A_{34}^T\hat B\end{array}\right]\left(I - \left[\begin{array}{cc}A_{22}^T & A_{32}^T \hat B\\ A_{23}^T & A_{33}^T\hat B
\end{array}\right]\right)^{-1}
\left[\begin{array}{c}A_{12}^T\\ A_{13}^T\end{array}\right]\tilde u_1^\#\\
&&+ A_{14}^T\tilde u_1^\#.
\end{eqnarray*}
Also, the corresponding matrices in the $(S,L,H)$ description of this system can be constructed using the formulas given in the proof of Theorem 1 in \cite{ShP5}. 


\end{document}